\documentclass[11pt,a4paper]{article}
\pdfoutput=1
\usepackage{jcappub}

\usepackage{cleveref}
\usepackage{amsmath,amsfonts,amssymb,mathrsfs,graphicx,color,longtable,bm}
\usepackage{hyperref}
\usepackage{graphicx}
\usepackage{array}
\usepackage{enumitem}
\usepackage[explicit]{titlesec}
\usepackage[utf8]{inputenc}
\usepackage[normalem]{ulem}
\usepackage{tablefootnote}

\hypersetup{
     colorlinks   = true,
     citecolor    = red,
     urlcolor     = blue,
     linkcolor    = blue
}

\usepackage{xcolor}
\usepackage[version=4]{mhchem}

\newcommand{\dd}{\mathrm{d}}

\makeatletter
\gdef\@fpheader{}
\makeatother

\begin{document}

\title{Knot reconstruction of the scalar primordial power spectrum with Planck, ACT, and SPT CMB data}

\author[a,b]{Antonio Raffaelli,}
\author[a,b,c,d]{Mario Ballardini}

\affiliation[a]{Dipartimento di Fisica e Scienze della Terra, Universit\`a degli Studi di Ferrara, via Giuseppe Saragat 1, 44122 Ferrara, Italy}
\affiliation[b]{INFN, Sezione di Ferrara, via Giuseppe Saragat 1, 44122 Ferrara, Italy}
\affiliation[c]{INAF/OAS Bologna, via Piero Gobetti 101, 40129 Bologna, Italy}
\affiliation[d]{Department of Physics \& Astronomy, University of the Western Cape, Cape Town 7535, South Africa}

\emailAdd{antonio.raffaelli@unife.it}
\emailAdd{mario.ballardini@unife.it}

\abstract{We investigate a non-parametric Bayesian method for reconstructing the primordial power spectrum (PPS) of scalar perturbations  using temperature and polarisation data from the {\em Planck}, ACT, and SPT CMB experiments. This reconstruction method is based on linear splines for the PPS between nodes in $k$-space whose amplitudes and positions are allowed to vary. 
All three data sets consistently show no significant deviations from a power-law form in the range $0.005 \lesssim k\,\mathrm{Mpc} \lesssim 0.16$ independent of the number of knots adopted to perform the reconstruction. The addition of high-resolution CMB measurements from ACT and SPT slightly improves the range of scales of the scalar PPS which are well constrained around a power law up to $k \simeq 0.25\,\mathrm{Mpc}^{-1}$ and $k \simeq 0.2\,\mathrm{Mpc}^{-1}$, respectively. At large scales, a potential oscillatory feature in the primordial power spectrum appears when we consider six or more nodes. 
We test the robustness of the methodology and our results by varying the detailed number of knots from $N=2$ to $N=10$.
We have used the reconstructed scalar PPS to derive several quantities related to inflationary dynamics, such as the effective scalar spectral index, which describes the dependence of the PPS on the scales and parameters associated with the effective field theory of inflation, to provide information on possible departures from the standard single-field canonical case. Finally, we investigate whether the excess of smoothing in the region of the acoustic peaks of the CMB anisotropy temperature power spectrum in the \textit{Planck} PR3 data is degenerate with our reconstructions of the PPS, but find no significant correlation between them.}

\maketitle

\section{Introduction}
Understanding the initial conditions of cosmic structure is a cornerstone of modern cosmology. The primordial power spectrum (PPS) of density perturbations, $\mathcal{P}_\mathcal{R}(k)$, encodes critical information about the physics of the early Universe, providing a direct link to inflationary dynamics and the underlying microphysical processes. The cosmic microwave background radiation (CMB) serves as a unique observational window into these initial conditions, as its anisotropies are sensitive to the properties of the PPS over a wide range of scales. 

The simplest inflation models predict an almost scale-invariant PPS, in agreement with current observations from \textit{Planck} \cite{Planck_inflation_2013,Planck_inflation_2015,Planck_inflation_2018} and other CMB experiments \cite{ACT_Aiola_2020,ACT_Choi_2020,SPT_Balkenhol_2023}. However, deviations from exact scale invariance - such as primordial features or any scale dependence in the PPS - could provide insights into the detailed mechanism of cosmic inflation, the particle content during the inflationary epoch, or even alternative theories of the early universe; see for reviews Refs.~\cite{Chluba_review,Achucarro_review}. Reconstructing the PPS directly from the data is therefore of fundamental importance.

Numerous approaches to the reconstruction of the PPS have been explored in the literature, broadly classified into parametric and non-parametric methods. Parametric methods are further subdivided into two main categories. In the first, the PPS is modelled using specific functional templates, such as power-law or broken power-law forms or more convolved shapes, with a finite set of parameters constrained using cosmological data; see for instance Refs.~\cite{Adams:2001vc,Peiris:2003ff,Covi:2006ci,Hamann:2007pa,Meerburg:2011gd,Planck_inflation_2013,Meerburg:2013dla,Benetti:2013cja,Miranda:2013wxa,Easther:2013kla,Chen:2014joa,Achucarro:2014msa,Hazra:2014goa,Hu:2014hra,Planck_inflation_2015,Gruppuso:2015zia,Gruppuso:2015xqa,Hazra:2016fkm,Torrado:2016sls,Planck_inflation_2018,Zeng:2018ufm,Canas-Herrera:2020mme,Braglia:2021ckn,Braglia:2021sun,Braglia:2021rej,Naik:2022mxn,Hamann:2021eyw,Antony:2024vrx}. In the second category, methods such as principal component analysis (PCA)~\cite{Leach:2005av,Dvorkin:2010dn} or deconvolution techniques, such as the modified Richardson-Lucy algorithm~\cite{Hazra:2013eva,Hazra:2014jwa,Sohn:2022jsm} and the Tikhonov regularisation~\cite{Hunt:2013bha}, introduce a finite number of degrees of freedom by selecting a reduced basis or regularised representation for the PPS. While these approaches do not assume a specific functional form, the choice of basis or regularisation introduces an implicit parameterisation, which places them within the parametric class.
On the other hand, non-parametric methods make minimal assumptions about the shape of the PPS and reconstruct it directly from data without imposing strong prior constraints. Techniques such as regularised deconvolution, Gaussian process regression, and moving knots reconstruction fall within this category (e.g., Refs.~\cite{Wang:2000js,Bridle:2003sa,Mukherjee:2003ag,Gauthier:2012aq,Vazquez_2012,Planck_inflation_2013,Aslanyan_2014,Abazajian_2014,Planck_inflation_2015,Finelli_2018,Obied:2017tpd,Planck_inflation_2018,Handley:2019fll,Chaki:2025qsc}). These approaches have revealed intriguing indications of primordial features in the PPS, although such findings remain statistically marginal.

In this paper, we focus on a non-parametric reconstruction of the PPS of scalar fluctuations using a moving node approach inspired by the method described in Ref.~\cite{Handley:2019fll}. This method uses a flexible representation of the PPS by a set of adaptive basis functions (or \textit{knots}) whose positions and amplitudes are optimised to fit the data. By using a Bayesian framework, we account for both statistical uncertainties and systematic effects, ensuring a robust and unbiased reconstruction. Compared to other non-parametric methods, the moving knots technique offers high fidelity in capturing potential features in the primordial fluctuations while maintaining computational efficiency.
In addition, methods that rely heavily on parametric assumptions often risk overfitting the intrinsic noise or amplifying scattering present in the data. The moving knots reconstruction approach, as adopted in this work, inherently addresses these challenges by offering a flexible, yet robust, framework for fitting the data without introducing spurious features. This method allows for a smooth interpolation between data points, ensuring the reconstructed spectrum faithfully represents the underlying physics rather than artefacts of the measurement process. Similar concerns have been discussed and mitigated in related works as in Refs.~\cite{Gauthier:2012aq,Planck_inflation_2013,Planck_inflation_2015,Planck_inflation_2018,Handley:2019fll}.
This technique, dubbed as \texttt{flexknot} after Ref.~\cite{Millea:2018bko}, has been successfully applied also in the reconstruction the evolution of the dark energy equation of state \cite{AlbertoVazquez:2012ofj,Hee:2015eba,Ormondroyd:2025exu}, the history of cosmic reionization \cite{Millea:2018bko,Heimersheim:2021meu}, galaxy cluster profiles \cite{Olamaie:2017vtt}, and the 21 cm signal \cite{Heimersheim:2023zyi,10.1093/mnras/stae614}.

We apply this reconstruction to the most recent CMB data, focusing on its implications for inflationary physics and possible deviations from the standard power-law initial conditions assumed within the $\Lambda$CDM model. We analyse \textit{Planck} data release PR3 and PR4, Atacama Cosmology Telescope DR4 (ACTPol), and South Pole Telescope (SPT-3G). Our results not only confirm the consistency of the PPS of scalar perturbations with near-scale invariance, but also explore the presence of features that could be associated with specific inflationary scenarios.

The paper is structured as follows. In~\cref{sec:theory}, we review the theoretical framework and the main quantities that we will reconstruct from the reconstructed PPS. We describe the moving knot reconstruction method in~\cref{sec:knots}. In~\cref{sec:data}, we introduce the cosmological data sets used in the analysis. We show and discuss the results of our analysis in~\cref{sec:results}. Finally, we summarise our results and outline prospects for future work in~\cref{sec:conclusions}. 
In the appendices we present further analyses performed to check the robustness of our analysis: in
\cref{app:parameters} we show the mean values of the cosmological parameters for each run, in \cref{app:npipe} we compare the reconstruction done with \textit{Planck} PR3 and \textit{Planck} PR4 data sets, and in \cref{app:alens} we study the impact of the parameter $A_\mathrm{L}$ on the reconstruction.

\section{Inflationary basics} \label{sec:theory}
Cosmic inflation \cite{Starobinsky_1980,Guth_1981,Linde_1982,Albrecht_Steinhardt_1982,Hawking_Moss_Stewart_1982,Linde_1983} is a theoretical framework that describes a phase of accelerated expansion in the early universe. This paradigm not only addresses key problems in standard Big Bang cosmology, such as the horizon, flatness, and monopole problems, but also provides a natural mechanism for the generation of primordial perturbations. These perturbations, which originate as quantum fluctuations in the {\em inflaton} field during the inflationary epoch, are stretched to macroscopic scales by the rapid expansion of the universe. After the end of inflation, these fluctuations serve as seeds for the formation of cosmic structure, manifesting as anisotropies in the CMB and as the inhomogeneities observed in the large-scale structure (LSS) of the Universe. By studying these cosmological observables, we can probe the physics of the very early universe and gain insights into inflationary dynamics shaping the initial conditions of the universe.

\subsection{Background equations}
In the simplest models of inflation, perturbations are seeded by density fluctuations of the inflaton field at horizon crossing scales. 
In single field models of inflation the dynamics of the inflaton $\phi$ with potential $V(\phi)$ is described by the action 
\begin{equation}
    S=\int \dd^4x\sqrt{-g}\left[\frac{M_\text{Pl}^2R}{2}-\frac{1}{2}\partial_\mu\phi\partial^\mu\phi-V(\phi)\right] \,,
    \label{eq:inflaton_action}
\end{equation}
where $M_\text{Pl}^2 \equiv 1/(8\pi G)$ with $G$ the gravitational constant, and $R$ is the Ricci scalar. The background metric is the Friedmann-Lemaitre-Robertson-Walker (FLRW) one 
\begin{equation}
    \dd s^2 = a^2(t)\left[-\dd \tau^2 + \dd\mathbf{x}^2\right] \,,
    \label{eq:FLRW_metric}
\end{equation}
where $\tau$ is conformal time defined as $\dd \tau \equiv a(t) \dd t$ with $t$ proper time. From the solution of the Einstein equations one gets the Friedmann and the Klein-Gordon equations that describe the evolution of the background dynamics. 
Assuming that the energy density of the Universe is dominated by the inflaton, we obtain
\begin{subequations}
    \begin{align}
        &H^2 = \frac{1}{3M_\text{Pl}^2}\left(\frac{\dot{\phi}}{2} + V\right) \,,\\
        &\ddot{\phi} + 3H\dot{\phi} + V_\phi = 0 \label{eq:inflaton_eom} \,,
    \end{align}
\end{subequations}
where $H \equiv \dot{a}/a$ is the Hubble parameter, $V_\phi$ is the derivative of the potential with respect to $\phi$ and a dot $\dot{\ }$ denotes derivative with respect to proper time $t$.

It is possible to parametrise the background dynamics of the inflationary universe through different parameters \cite{Starobinsky:1979ty,Mukhanov:1985rz,Mukhanov:1988jd,Stewart:1993bc,Liddle:1994dx,Gong:2001he,Schwarz:2001vv,Leach:2002ar,Vennin:2014xta,Auclair:2022yxs,Bianchi:2024qyp,Ballardini:2024irx}. 
For instance, we can use the Hubble-flow functions (HFFs) defined as 
\begin{equation}
        \epsilon_1 \equiv \epsilon = -\frac{\dot{H}}{H^2} \,, \quad
        \epsilon_2 \equiv \eta = \frac{\dot{\epsilon}}{H\epsilon} \,, \quad
        \epsilon_n = \frac{\dd \ln \left|\epsilon_{n-1}\right|}{\dd N} \,,
    \label{eq:slow_roll_parameters}
\end{equation}
where $N$ is the number of $e$-folds and it is defined as $\dd N \equiv H \dd t$. The slow-roll inflationary dynamics, i.e. the inflationary phase long enough to solve the flatness and horizon problems, is guaranteed by $\epsilon_n \ll 1$. In general, in single-field slow-roll inflation it is also possible to describe the primordial scalar and tensor fluctuations in terms of HFFs, resulting in a unified framework for linking the predictions of hundreds of slow-roll inflationary models to cosmological observations; see Refs.~\cite{Martin:2013tda,Martin:2024qnn,Ballardini:2024ado}.

\subsection{Primordial power spectrum from quantum fluctuations}
One can define quantum fluctuations of the inflaton field as small perturbations around a mean field as
\begin{equation}
    \phi(t,x) = \Bar{\phi}(t) + \delta\phi(t,x) \,.
    \label{eq:inflaton_quantum_fluctuation}
\end{equation}

Information on the distribution of quantum fluctuations from inflation is encoded in their power spectrum, that is the Fourier transform of the correlation function of the fluctuations, namely 
\begin{equation}
    \langle\delta\phi(\mathbf{k},\tau)\delta\phi(\mathbf{k}',\tau)\rangle = \left(2\pi\right)^3\delta^{(3)}(\mathbf{k}-\mathbf{k}') P_{\delta\phi}(k) \,.
    \label{eq:inflaton_power_spectrum}
\end{equation}
where $\delta^{(3)}$ is the three-dimensional Dirac delta function.
One can then move from the fluctuations of the field to its density perturbations. For instance, in single-field slow-roll inflationary models one can show that 
\begin{equation}
    \frac{\delta\phi}{\dot{\phi}}\simeq\frac{\delta\rho}{\dot{\rho}} \,.
    \label{eq:delta_to_rho_inflation}
\end{equation}

At this point quantum fluctuations from inflation are related to the metric perturbations by the gauge-invariant curvature perturbation $\mathcal{R}$, since $\delta \phi$ is not. In particular, if we define the scalar perturbed metric as \cite{Bardeen_1980,Bardeen_1983,Lyth_1985,Deruelle_1995,Martin_1998,Malik_2009}
\begin{equation}
    \dd s^2 = a^2(\tau) \left[\left(1+2\Phi\right) \dd\tau^2 + \left(1-2\Psi\right) \dd\mathbf{x}^2\right]\,,
\end{equation}
where
\begin{equation}
    \mathcal{R}\equiv \Psi -\frac{\mathcal{H}\left(\mathcal{H}\Phi+\Psi^\prime\right)}{\mathcal{H}^\prime-\mathcal{H}^2} \,,
\end{equation}
and a prime $'$ denotes derivative with respect to conformal time $\tau$.
In Fourier space, the equation of motion of the gauge-invariant curvature perturbation is given by the Mukhanov-Sasaki equation \cite{Mukhanov_CHibisov_1981,Mukhanov:1985rz,Sasaki_1986,MUKHANOV1992203,Kodama_1984} 
\begin{equation}
    u^{\prime\prime}_k+\left(c_\mathrm{s}^2k^2-\frac{z^{\prime\prime}}{z}\right) u_k=0\,,
    \label{eq:mukhanov_sasaki}
\end{equation}
where $u_k \equiv z\mathcal{R}(k)$, $z = a\sqrt{2\epsilon_1}/c_\textrm{s}$, and $c_\textrm{s}$ is the speed of sound during inflation. Vacuum state solution is commonly chosen to be the Bunch-Davies vacuum \cite{Bunch:1978yq} corresponding to 
\begin{equation}
    u_k(\tau) = \frac{e^{-ik\tau}}{\sqrt{2k}}\left(1-\frac{i}{k\tau}\right) \,.
\end{equation} 
Information on the distribution of curvature on different scales is contained at first order in the primordial curvature power spectrum 
\begin{equation}
    \langle\mathcal{R}_\mathbf{k}(\tau)\mathcal{R}_\mathbf{k'}(\tau)\rangle = (2\pi)^3\delta^{(3)}\left(\mathbf{k}-\mathbf{k'}\right)P_\mathcal{R}(k)\,,
    \label{eq:two_point_PPS}
\end{equation}
and in general it is useful to define the dimensionless power spectrum as 
\begin{equation}
    \mathcal{P}_\mathcal{R}(k) \equiv \frac{k^3}{2\pi^2}P_\mathcal{R}(k) \,.
    \label{eq:dimensionless_PPS}
\end{equation}
Solving the equation of motion of the primordial perturbations in quasi de-Sitter slow-roll single-field inflation one can write the PPS on super-horizon scales as 
\begin{equation}
    \mathcal{P}_\mathcal{R}(k) = A_\mathrm{s} \left(\frac{k}{k_*}\right)^{n_\mathrm{s} - 1} \,,
    \label{eq:power_law_PPS}
\end{equation}
where $A_\mathrm{s}$ and $n_\mathrm{s}$ are the amplitude and the spectral index for scalar perturbations and $k_*$ is a pivot scale. This is actually a first-order approximation of the super-horizon solution of~\cref{eq:mukhanov_sasaki}. The most general solution allows for deviations from this power-law spectrum parametrised by the runnings of the scalar spectral index, namely non-zero higher derivatives of the primordial power spectrum \cite{Kosowsky_1995}
\begin{equation}
    \mathcal{P}_\mathcal{R}(k) = A_\mathrm{s}\left(\frac{k}{k_*}\right)^{n_\mathrm{s} - 1 + \frac{\alpha_\text{s}}{2}\ln\left(\frac{k}{k_*}\right) + \frac{\beta_\text{s}}{3!}\ln\left(\frac{k}{k_*}\right)^2 + \dots}
    \label{eq:power_law_PPS_full}
\end{equation}
where these phenomenological parameters can be perturbatively written in terms of HFFs.

\subsection{Slow-roll parameters reconstruction from EFT of inflation}
A more general approach to the primordial curvature perturbations starts with the definition of the Effective Field Theory (EFT) of inflation \cite{Cheung_2008,Palma_2015,Palma_2016,Durakovic:2019kqq}. In this framework one can write the most general action starting from operators that depend on the metric perturbations. In this context one can use the Arnowitt--Deser--Misner (ADM) formalism \cite{Arnowitt_2008} to write the inflation action in terms of the primordial curvature perturbation $\mathcal{R}$. 

In particular one gets the second order action as \cite{Ach_carro_2013,Palma_2015,Palma_2016} 
\begin{equation}
    S_2=\frac{1}{2}\int \dd^3x \dd\tau\left[\left(u'\right)^2-c_\mathrm{s}^2\left(\nabla u\right)^2+\frac{z''}{z}u^2\right] \,.
\end{equation}
Writing $c_\mathrm{s}^2 = 1-\theta$ one can separate the \textit{free action} from the action due to variations in $c_\mathrm{s}$ and other slow-roll parameters as
\begin{align}
    &S_0=\frac{1}{2}\int \dd x^3 \dd\tau\,\left[\left(u^\prime\right)-\left(\nabla u\right)^2+\frac{z_0^{\prime\prime}}{z_0}u^2\right]\,,\\
    &S_\text{int}^{(2)}=\frac{1}{2}\int \dd x^3 \dd\tau\left[\theta\left(\nabla u \right)^2+\frac{1}{\tau^2}\delta(\tau)u^2\right]
\end{align}
where the definition of $z_0$ and $\delta(\tau)$ comes from the first order expansion of $z^{\prime\prime}/z$
\begin{equation}
    \frac{z^{\prime\prime}}{z}\simeq\frac{1}{\tau^2}\left(2+3\epsilon+\frac{3\eta}{2}-3s-\tau\frac{\eta^\prime}{2}+\tau s^\prime\right)\equiv\frac{z_0^{\prime\prime}}{z_0}\left[1+\frac{1}{2}\delta(\tau)\right]\,, \qquad \frac{z_0^{\prime\prime}}{z_0}=\frac{2}{\tau^2} \,.
\end{equation}
Here $s$ is defined as $s \equiv c_\mathrm{s}^\prime/(aHc_\mathrm{s})$; we have expressed $\delta(t)=\delta_H-\tau\theta^\prime+\tau^2\theta^{\prime\prime}/2$ where $\delta_H=3\epsilon+3\eta/2-\tau\eta'/2$ includes only variations of the slow roll parameters in such a way to separate the part depending on the speed of sound during inflation. We recall that in the context of the EFT of inflation, $c_\mathrm{s}$ is an effective parameter that captures modifications to the dynamics of scalar perturbations arising from non-canonical interactions or additional degrees of freedom beyond a simple single-field slow-roll scenario.

In the context of \textit{in-in} formalism one can write the two-point function as 
\begin{equation}
\begin{aligned}
    \langle u(x, \tau) u(y, \tau) \rangle =\, &\langle 0 | u_I(x, \tau) u_I(y, \tau) | 0 \rangle 
+ \\
& +i \int_{\tau}^{-\infty} \dd\tau' \, \langle 0 | 
\big[ H_I^{(2)}(\tau'), u_I(x, \tau) u_I(y, \tau) \big] 
| 0 \rangle \,,
\end{aligned}
\end{equation}
and in our case the interaction Hamiltonian in the interaction picture is 
\begin{equation}
    H_I^{(2)}(\tau)=-\frac{1}{2}\int \dd^3x\left[\theta(\tau)\left(\nabla u_I\right)^2+\frac{\delta(\tau)}{\tau^2}u_I^2\right] \,.
\end{equation}
Defining deviations from \cref{eq:power_law_PPS} as
\begin{equation}
    \mathcal{P}_\mathcal{R}(k)=\mathcal{P}_0(k)+\Delta\mathcal{P}(k)\,,
\end{equation}
where $\mathcal{P}_0$ is the zero-order contribution to the power-law PPS. One can finally write $\Delta\mathcal{P}(k)$ in terms of variations of $\theta$ and $\delta_H$ as
\begin{equation} \label{eqn:deltaP}
    \frac{\Delta \mathcal{P}}{\mathcal{P}_0}(k) = k \int_{-\infty}^0 \dd\tau 
\left[
-\theta + \delta_H \frac{1}{k^2 \tau^2} + 2 \delta_H \frac{1}{k^4 \tau^4} 
- \frac{1}{k^4 \tau^3} \frac{\dd \delta_H}{\dd\tau}
\right] 
\sin(2k\tau) \,.
\end{equation}
Finally, writing the sine in its exponential form and imposing that both $\theta(\tau)$ and $\delta_H$ are odd functions of time it is possible to extend the integration domain such that \cite{Palma_2015}
\begin{equation}
    k^3\frac{\Delta\mathcal{P}}{\mathcal{P}_0}(k)=-\frac{1}{4}\int_{-\infty}^{+\infty}d\tau\,\left[\frac{1}{8}\theta^{\prime\prime\prime\prime}+\frac{\delta_H^{\prime\prime}}{2   \tau^2}-\frac{\delta_H}{\tau^4}\right]\sin(2k\tau) \,.
\end{equation}
At this point it is possible to Fourier invert this equation in order to derive $\theta$ or $\delta_H$, as functions of time, depending on $\Delta\mathcal{P}/\mathcal{P}_0$, as done in Refs.~\cite{Palma_2015,Palma_2016,Durakovic:2019kqq}.

\subsubsection{Variations in the sound speed}
First we set $\delta_H=0$ to study the effect of variations in $\theta(\tau)$ alone. After some integration by parts and Fourier inversion, the result for $\theta$ is
\begin{equation} \label{eqn:theta}
    \theta(\tau)=-\frac{4}{\pi}\int _0^{+\infty}\frac{\dd k}{k}\frac{\Delta\mathcal{P}}{\mathcal{P}_0}(k)\sin(2k\tau)\,.
\end{equation}
One should keep in mind that this results works in the assumption that $\theta$ alone inherits all the features in the PPS \cite{Achucarro:2010da,Chen:2011zf}.
\Cref{eqn:theta} matches the one previously obtained in Refs.~\cite{Ach_carro_2013,Palma_2015}.

\subsubsection{Variations of the background}
We now consider the case where only the background quantities, namely $\delta_H$, vary, fixing $c_\mathrm{s}$ to unity. After inversion we find 
\begin{equation}
    \frac{\delta_H^{\prime\prime}}{2\tau^2}-\frac{\delta_H}{\tau^4}=\frac{4}{\pi i}\int \dd k\,  k^3\frac{\Delta\mathcal{P}}{\mathcal{P}_0}(k)e^{-2ik\tau}\,.
    \label{eq:slow_roll_equation}
\end{equation}
This differential equation can be solved through Green's function method. In particular solving 
\begin{equation}
    \frac{g^{\prime\prime}(k,\tau)}{2\tau^2}-\frac{g(k,\tau)}{\tau^4}=e^{-2ik\tau}
    \label{eq:green_epsilon}
\end{equation}
one can obtain the following solution
\begin{equation}
    \delta_H(\tau)=\frac{4}{\pi i }\int \dd k\,k^3\frac{\Delta\mathcal{P}}{\mathcal{P}_0}(k)g(k,\tau)\,,
    \label{eq:delta_H_green}
\end{equation}
where the Green's function $g(k,\tau)$ is found solving \cref{eq:green_epsilon} and fixing the integration constants by imposing that the slow-roll parameters are odd functions of time. One gets
\begin{equation}
    g(k,\tau)=-\frac{\tau^2}{6k^2}+\frac{i}{2k^5\tau}+e^{-2ik\tau}\left(-\frac{i}{2k^5\tau}-\frac{\tau^2}{2k^2}+\frac{1}{k^4}+\frac{i\tau}{k^3}\right)\,,
\end{equation}
and substituting in~\cref{eq:delta_H_green} one obtains the final solution for slow-roll parameters as 
\begin{equation} \label{eqn:deltaH}
    \delta_H(\tau)=\frac{4}{\pi}\int_0^{+\infty}\frac{\dd k}{k}\frac{\Delta\mathcal{P}}{\mathcal{P}_0}(k)\left[\frac{1}{k\tau}+\left(k^2\tau^2-2\right)\sin(2k\tau)+\left(2k\tau-\frac{1}{k\tau}\right)\cos(2k\tau)\right] \,.
\end{equation}

\subsubsection{Sudden variations of the background}
Similarly to the previous subsection, we consider also here the component exclusively due to variations in the background parameters. 
However, we are interested in sudden variations associated with features in the PPS \cite{Starobinsky:1992ts,Adams:2001vc,Chen:2006xjb,Chen:2008wn,2025JCAP...04..007C}, which means that time-derivative terms will dominate, leading to $\delta_H \simeq -\tau\eta'/2$. Under this assumption, and after Fourier inversion, \cref{eq:slow_roll_equation} becomes
\begin{equation}
    -\frac{1}{4\tau}\eta^{\prime\prime\prime}=\frac{4}{\pi i}\int \dd k\, k^3\frac{\Delta\mathcal{P}}{\mathcal{P}_0}(k)e^{-2ik\tau}\,,
\end{equation}
and the corresponding Green's function is 
\begin{equation}
    g(k,\tau)=-\frac{3}{4k^4}+\frac{i}{k^3}\tau+\frac{1}{2k^2}\tau^2+\frac{e^{-2ik\tau}}{2k^3}\left(\frac{3}{4k}+i\tau\right)\,.
\end{equation}
Finally, one gets
\begin{equation} \label{eqn:eta}
    \eta(\tau)=\frac{4}{\pi}\int_0^{+\infty} \dd k\, \frac{\Delta\mathcal{P}}{\mathcal{P}_0}(k)\left[2\tau -\frac{3}{2k}\sin(2k\tau)+\tau\cos(2k\tau)\right]\,.
\end{equation}

\section{Primordial power spectrum reconstruction formalism} \label{sec:knots}
\begin{figure}
    \centering
    \includegraphics[width=0.5\linewidth]{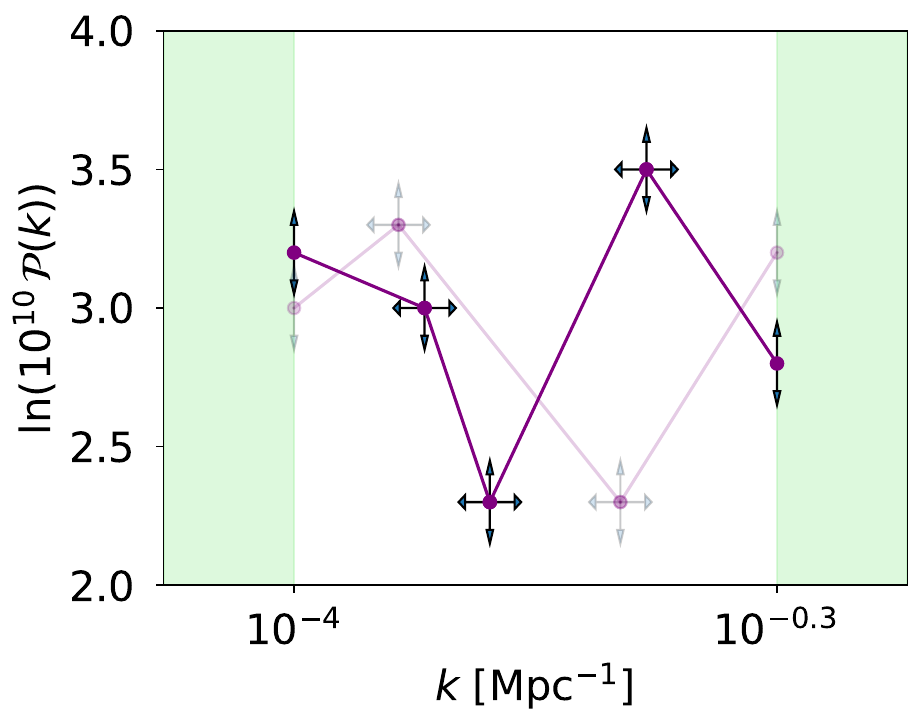}
    \caption{Visualisation of the knot reconstruction. Points' coordinates in the spectrum are sampled and the spectrum results from linear interpolation of the points.}
    \label{fig:knots}
\end{figure}

Knot reconstruction of the scalar PPS is a model-agnostic way of looking for features in the initial conditions of the Universe and for this reason it has already been employed largely in the literature \cite{Vazquez_2012,Aslanyan_2014,Ravenni_2016,Planck_inflation_2018}. 
Knots are points defined by the wavenumber $k$ and the amplitude of the PPS $\mathcal{P}_{\mathcal{R}}$ as shown in~\cref{fig:knots}.
Fixing the number of knots and varying the position $k$ and the amplitude $\mathcal{P}_{\mathcal{R}}$ for each knot allows to explore the whole parameter space and look for non-trivial features. 

Although this approach is model independent, the reconstruction in general depends on the choice of the method used. The most immediate example is the freedom of choice among the interpolating function of the knots. In this work we chose a linear interpolation, but in the literature different choices have also been made, such as cubic interpolation \cite{Hu_2014,Handley:2019fll}. In addition, the number of knots is an arbitrary degree of freedom.

As our baseline, the PPS is computed as a linear interpolation of the knots. Specifically, we implement the following algorithm
\begin{equation} \label{eq:spline_interpolation}
    \ln\left(10^{10}\mathcal{P}_\mathcal{R}(k)\right)=\sum_{i=0}^{N-1}\left[\frac{y_{i+1}-y_i}{k_{i+1}-k_i}\left(k-k_i\right)+y_i\right] \left[\Theta(k-k_i)-\Theta(k-k_{i+1})\right]\,,
\end{equation}
where $y_i = \ln\left(10^{10}\mathcal{P}_\mathcal{R}(k_{i})\right) \equiv \ln\left(10^{10}\mathcal{P}_i\right)$ and $\Theta(x)$ is the Heaviside function. 
One thing to note is that the nodes must be sorted, i.e. ordered so that $k_i<k_{i+1}$ for any node $i$, otherwise the reconstructed spectrum would be unphysical. This is done through the \textit{forced identifiability prior} already implemented in {\tt Polychord} \cite{Polychord1_Handley_2015,Polychord2_Handley_2015},\footnote{\url{https://github.com/PolyChord/PolyChordLite}} which allows parameters to be sorted without spoiling the computation of the Bayesian evidence for the sampled parameters.

\section{Cosmological data sets} \label{sec:data}
In this section, we describe the data sets used for the analysis. In this work we have focused on measurements from CMB experiments in order to reconstruct the scalar PPS at different scales. As mentioned in the previous section, we use \texttt{PolyChord} as a sampler, and for each run we set the number of live points to 1000. 

Together with \texttt{PolyChord} we use {\tt Cobaya} \cite{Cobaya_Torrado_2021}.\footnote{\url{https://github.com/CobayaSampler/cobaya}}
We split the analysis into two different ways of sampling the positions $k$: in one case we sample $k$ with a uniform prior between $[10^{-4},10^{-0.3}]$ $\text{Mpc}^{-1}$, in the other we sample in the same range but with a log-uniform prior.  
These different choices allow us to probe the whole range of scales: in the latter case, the search for features focuses on large scales, while in the former case the different sampling allows us to focus on the small scales.

The choice of the range $k\in[10^{-4},10^{-0.3}]$ $\text{Mpc}^{-1}$ is related to the projected scales probed by the CMB experiments. It has been tailored to recover the results assuming a scalar power-law PPS; in particular in the case with two knots. These assumptions on the range are the same as those made in Ref.~\cite{Handley:2019fll}. 

We use a modified version of {\tt CAMB} \cite{CAMB_Lewis_2000,Howlett_2012}.\footnote{\url{https://github.com/cmbant/CAMB}} For each run we fix $N_\text{eff} = 3.044$ \cite{Akita_2020,Froustey_2020,Bennett_2021} and a massive neutrino with $m_\nu=0.06$ eV.
For the calculation of the nonlinear matter power spectrum affecting CMB lensing and lensed components, we use {\tt HMcode} \cite{Mead_2021}.\footnote{\url{https://github.com/alexander-mead/HMcode}} 
We fix the primordial \ce{^{4}He} mass fraction $Y_{\rm p}$, taking into account the different values of the baryon fraction $\omega_{\rm b}$, tabulated in the public code {\tt PArthENoPE} \cite{Pisanti:2007hk,Consiglio:2017pot,Gariazzo:2021iiu}. 
All priors for the parameters are listed in \cref{tab:priors}.

The data sets used in our analysis are the following:
\begin{itemize}
    \item \textbf{Planck 2018}. We use the \textit{Planck} PR3 data sets and likelihood codes \cite{Planck_likelihood_2020}. At low multipoles we use the \texttt{Commander} likelihood in temperature and the \texttt{SimAll} likelihood for large-scale $E$-mode polarization. At high multipoles we use the \texttt{plik\_lite} \textit{Planck} likelihood for $TT$, $TE$, $EE$ covering the multipole range $30 \leq \ell \leq 2508$ where the standard \texttt{Plik} likelihood foreground and nuisance parameters are marginalised. Finally, we consider the \textit{Planck} 2018 lensing likelihood \cite{Planck_2018_lensing} over the ``conservative'' multipole range $8 \leq L \leq 400$. We will refer to the combination of all these data sets as P18.
    \item \textbf{Planck NPIPE + SRoll2}. We consider the high-multipole likelihood based on \textit{Planck} PR4 maps (\texttt{NPIPE}) \cite{Planck:2020olo} using the \texttt{CamSpec} likelihood \cite{camspec_npipe_2022,Efstathiou_2021}. At low multipoles we consider the \texttt{Commander} likelihood in temperature and for polarisation the \texttt{SimAll} likelihood based on \texttt{SRoll2} products \cite{Delouis_2019,Pagano:2019tci}.
    We use the \textit{Planck} PR4 lensing likelihood based on \texttt{NPIPE} maps \cite{Carron_2022}. We will refer to the combination of these data sets as NPIPE.
    \item \textbf{ACTPol}. We use the \texttt{pyactlike} likelihood for ACTPol DR4 data \cite{ACT_Aiola_2020,ACT_Choi_2020}.\footnote{\url{https://github.com/ACTCollaboration/pyactlike}} In combination with \textit{Planck} we use the standard likelihood with $TT$ data starting from $\ell=1500$ to avoid for unaccounted correlation between the two data sets while $TE$ and $EE$ data always starts from $\ell=350$. The likelihood data range up to $\ell_\mathrm{max} = 4125$. In the following we will also examine the effect of removing the last data points, specifically using data up to $\ell_\mathrm{max} \leq 2800$.  We will refer to these data as ACTPol when there is no cut in the data points and ACTPol-cut when the last data points are removed.
    \item \textbf{SPT-3G}. We use the SPT-3G likelihood described in Ref.~\cite{SPT_Balkenhol_2023}.\footnote{\url{https://github.com/xgarrido/spt_likelihoods}} This data set covers the multipole range $750 < \ell < 3000$ for $TT$ and $300< \ell < 3000$ for $TE$ and $EE$. For the nuisance parameters we used the standard prior ranges, see Ref.~\cite{SPT_Balkenhol_2023}. We will refer to these data as SPT-3G.
\end{itemize}

\begin{table}[t]    
    \centering
    \begin{tabular} { l  c c }
    \noalign{\vskip 3pt}\hline\noalign{\vskip 1.5pt}\hline\noalign{\vskip 5pt}
    \multicolumn{1}{c}{\bf Parameter} &   \multicolumn{1}{c}{\bf Prior type} &  \multicolumn{1}{c}{\bf Prior range}\\
    \noalign{\vskip 3pt}\cline{1-3}\noalign{\vskip 3pt}
    $\Omega_\mathrm{c}h^2$  & Uniform  & [0.095,\,0.145] \\
    $\Omega_\mathrm{b}h^2$  &  Uniform & [0.019,\,0.025] \\
    $\tau_\text{reio}$  & Uniform  & [0.01,\,0.4] \\
    $100\,\theta_\text{MC}$  & Uniform  & [1.03,\,1.05] \\
    $k_i$ [Mpc$^{-1}$]  & Sorted Log-Uniform  &  $[10^{-4},\,10^{-0.3}]$\\
    $\ln\left(10^{10}\,\mathcal{P}_i\right)$  &  Uniform & [2,\,4]\\\hline
    \noalign{\vskip 1.5pt}
    $\ln\left(10^{10}A_\mathrm{s}\right)$\tablefootnote{$A_\mathrm{s}$ and $n_\mathrm{s}$ are only used when considering a power-law PPS within the $\Lambda$CDM model.}  & Uniform  & [1.61,\,3.91]\\
    $n_\mathrm{s}$  & Uniform  &  [0.8,\,1.2]\\
    \hline\noalign{\vskip 1.5pt}\hline
    \end{tabular}
    \caption{Priors on the cosmological parameters.}
    \label{tab:priors}
\end{table}

\section{Results} \label{sec:results}
In this section, we present the results of our analysis using different combinations of the data sets described in the previous section: P18, P18 + ACTPol, and P18 + SPT-3G. We reconstruct the scalar PPS using the moving knot approach, varying the number of knots $N$ between 2, 4, 6, 8, and 10. The positions of the knots are sampled both uniformly and logarithmically to explore different resolution schemes. This allows us to assess the impact of data set selection and knot placement on reconstruction accuracy and stability.

\subsection{Reconstrunction from CMB data}
\begin{figure}
    \centering
    \includegraphics[width=1\linewidth]{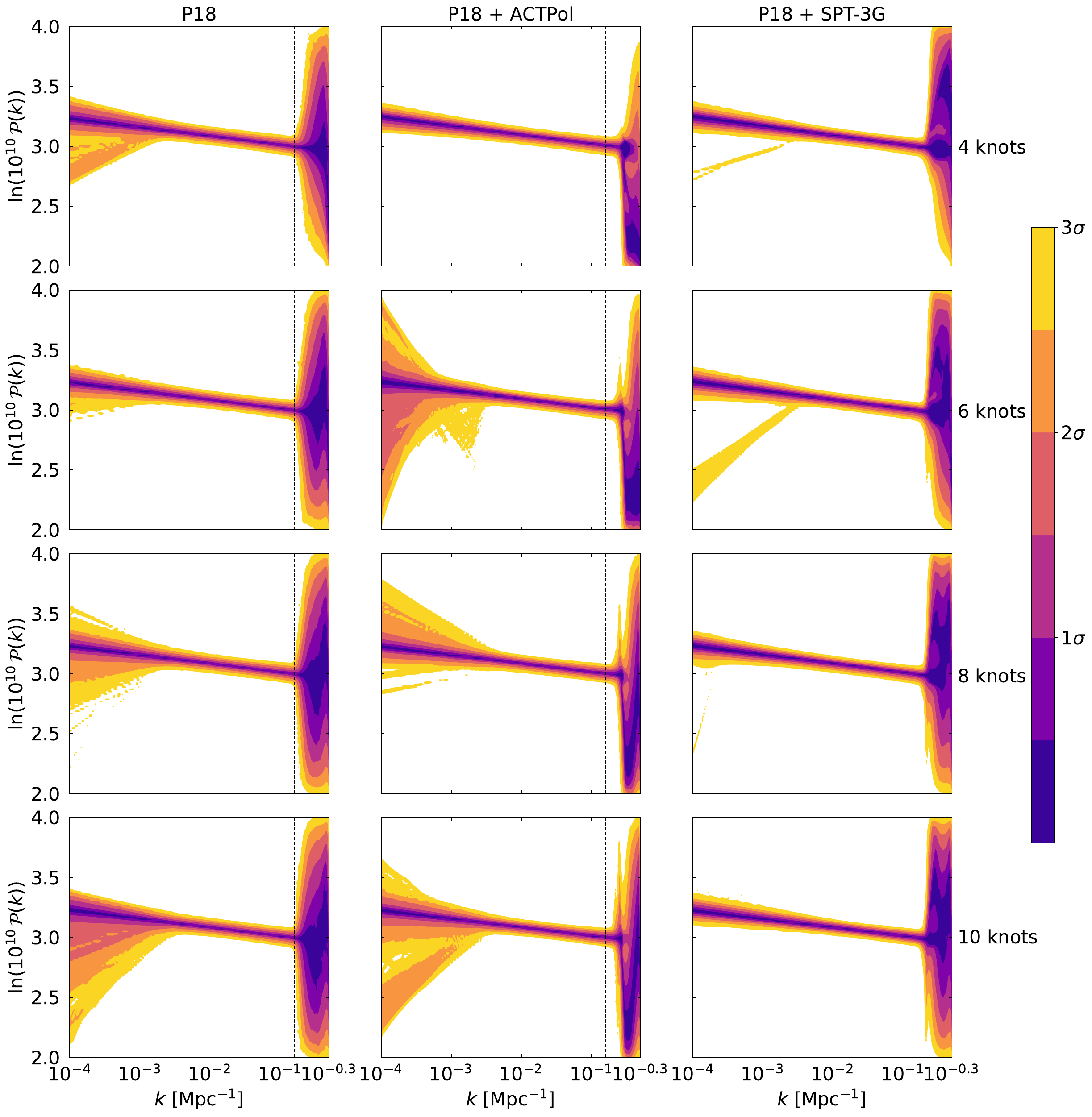}
    \caption{Reconstruction of the scalar PPS plotted using iso-probability credibility intervals with their masses converted to $\sigma$-values via an inverse error function transformation, obtained for each data set. We consider $N$ knots with positions $k_i$ (except for the outermost ones) uniformly sampled in the range $[10^{-4},10^{-0.3}]\, \mathrm{Mpc}^{-1}$.}
    \label{fig:spectra_lin}
\end{figure}

\begin{figure}[htp]
    \centering
    \includegraphics[width=1\linewidth]{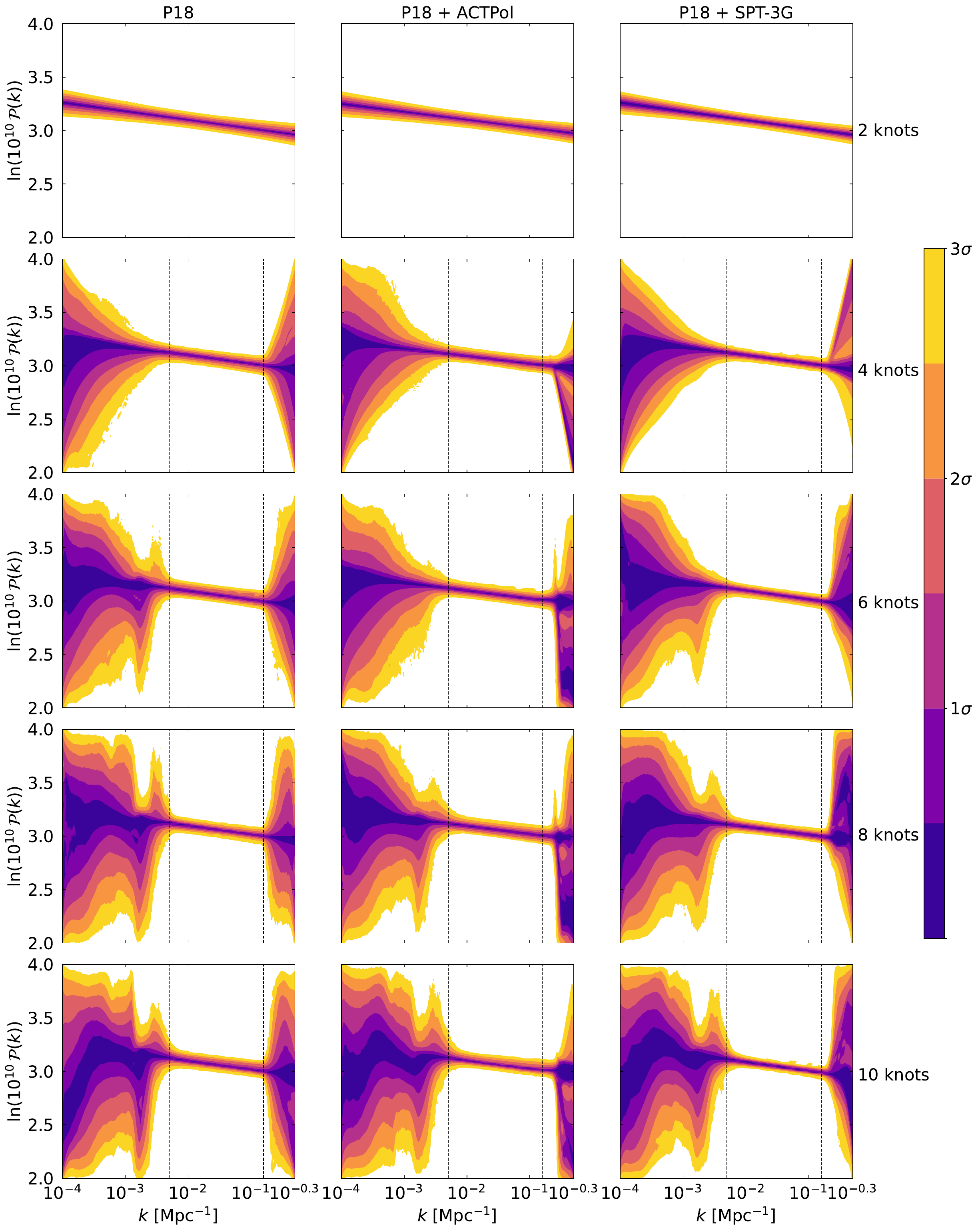}
    \caption{Same as \cref{fig:spectra_lin}, but with positions $k_i$ (except for the outermost ones) logarithmically sampled in the range $[10^{-4},10^{-0.3}]\, \mathrm{Mpc}^{-1}$.}
    \label{fig:spectra_log}
\end{figure}

\Cref{fig:spectra_lin} shows the results obtained sampling the positions of the knots with a uniform prior for different numbers of knots, considering the \textit{Planck} data set alone as well as its combination with ACTPol and SPT-3G, separately. 
\Cref{fig:spectra_log} shows the results obtained sampling the positions of the knots with a log-uniform prior for each data sets.
The posterior distributions of the reconstructed functions are visualised using the {\tt fgivenx} package \cite{fgivenx},\footnote{\url{https://github.com/handley-lab/fgivenx}} which provides a clear representation of the uncertainty in the reconstruction.

A first observation concerns the effect of the different sampling strategies on the reconstruction. Linear sampling places a greater emphasis on small scales, whereas logarithmic sampling provides better sensitivity to potential features at larger scales. This difference is crucial in identifying deviations from a simple power-law power spectrum over a wide range of wavenumbers $k$.

\begin{figure}
    \centering
    \includegraphics[width=1\linewidth]{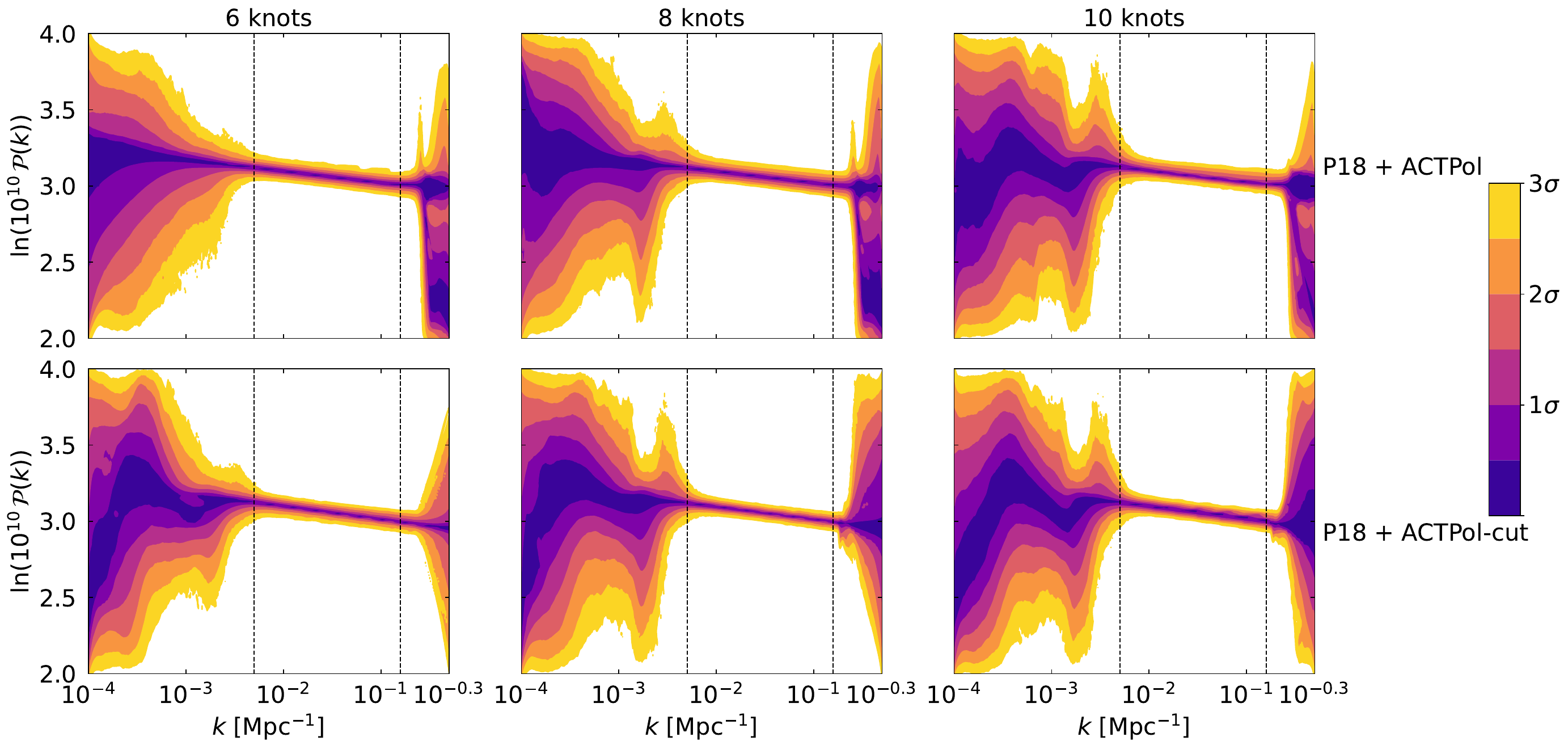}
    \caption{Same as \cref{fig:spectra_log}, we compare the reconstructions obtained with the combination P18 + ACTPol using the full ACTPol data set up to $\ell_\text{max}^\text{ACTPol} = 4125$ (upper panels) with those obtained with the combination P18 + ACTPol using the ACTPol data cut above $\ell_\text{max}^\text{ACTPol}=2800$ (lower panels).}
    \label{fig:act_meno4}
\end{figure}

Focusing on the logarithmic sampling case, we observe that increasing the number of knots leads to the emergence of oscillatory features at $k \sim 10^{-2.75}\, \mathrm{Mpc}^{-1}$ corresponding to the known trough on the CMB temperature angular power spectrum at $\ell \sim 25$. 
However, these oscillations remain consistent with a power-law spectrum within 68\% confidence level (CL). In contrast, the corresponding feature in the CMB angular power spectrum is much more pronounced and reaches a higher statistical significance. 
We conclude that P18 provides strong constraints on the scalar PPS, keeping it close to a power-law form in the range $0.005 \lesssim k\,\mathrm{Mpc} \lesssim 0.16$ where these numbers correspond to the vertical dashed lines plotted in~\cref{fig:spectra_lin,fig:spectra_log,fig:act_meno4,fig:npipe,fig:alens}.

On small scales, the addition of high-resolution CMB data sets, such as ACTPol or SPT-3G, does not significantly alter the reconstruction obtained using \textit{Planck} data alone, aside from a minor reduction in noise; in particular we observe that P18+ACTPol reconstruction shows power law behaviour up to $k\sim0.25$ Mpc$^{-1}$ and for P18+SPT-3G reconstruction up to $k\sim0.2$ Mpc$^{-1}$.  
This suggests that, at these scales, \textit{Planck} data already provide most of the constraining power on the scalar PPS, and the addition of current high-resolution measurements do not introduce qualitatively new features.

However, a notable exception is found when ACTPol data are included. The reconstruction with ACTPol shows a clear suppression of the power spectrum amplitude on small scales. This effect is even more pronounced in the case of linear sampling; see \cref{fig:spectra_lin}. In particular, for large $k$ the scalar PPS shows two best-fit regions, one of which deviates from the standard power-law solution. We interpret this as a consequence of the fact that some of the high-multipoles of the ACTPol data points take on negative mean values, albeit with large error bars \cite{ACT_Choi_2020,ACT_Aiola_2020}. This highlights the sensitivity of the knot-based reconstruction method in capturing and reflecting specific features in the data. 
To investigate this effect further, we repeat the analysis for the combination of \textit{Planck} and ACTPol data, removing the last four data points from ACTPol in $TT$, $TE$, and $EE$, effectively lowering the maximum multipole used in the analysis to $\ell_\text{max} =2800$; the results are shown in \cref{fig:act_meno4}. This cut eliminates the feature at small scales, confirming its dependence on the high-$\ell$ ACTPol measurements. Interestingly, suppressing this feature also seems to allow for a more pronounced trough at large scales in the reconstructed PPS.

In~\cref{app:npipe}, we compare the results obtained using \textit{Planck} PR3 with those based on PR4 \texttt{NPIPE} maps and \texttt{SRoll2} low-$\ell$ polarization. The reconstructed power spectra are fully consistent, as shown in~\cref{fig:npipe}.

\begin{figure}
    \centering
    \includegraphics[width=0.4\linewidth]{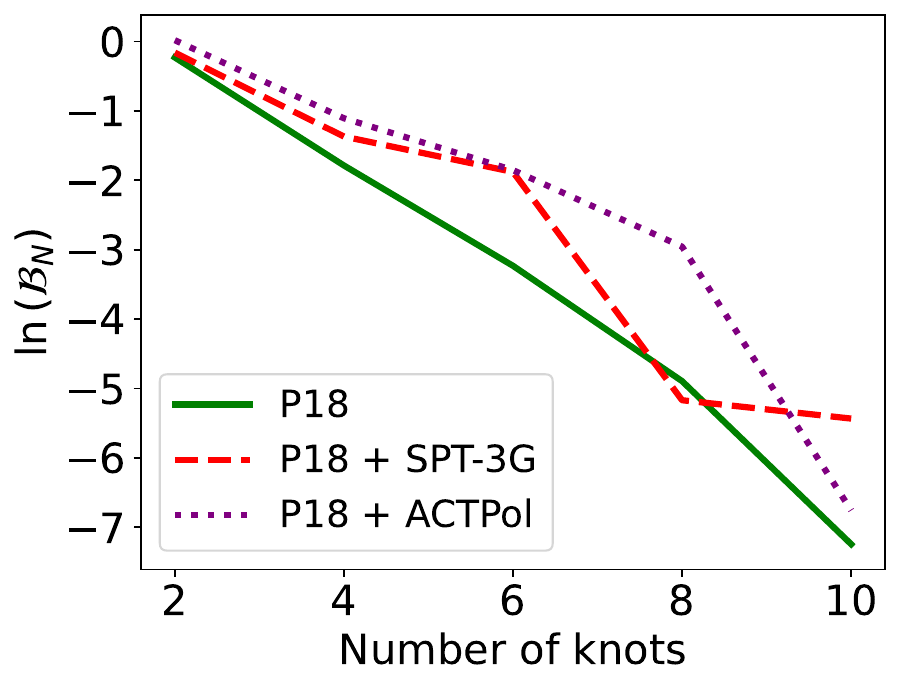}
    \caption{Bayes factors for logarithmic sampling reconstruction for the different data sets varying the number of knots.}
    \label{fig:evidence}
\end{figure}

In~\cref{fig:evidence}, we show the Bayes factor expressed with respect to the power-law case for each of the data sets defined as
\begin{equation}
    \ln\left(\mathcal{B}_N\right)=\ln \left(\frac{\mathcal{Z}_N}{\mathcal{Z}_\text{PL}}\right) \,.
\end{equation}
Results are shown varying the number of nodes $N$ for the logarithmic sampling reconstruction.
We observe no statistical preference over the power law or the two nodes case.

\subsection{Reconstruction of the scalar spectral index}
From the reconstructed scalar PPS, one can extract information about the effective scalar spectral index $n_\mathrm{s}$. This follows directly from the definition
\begin{equation} 
    n_\mathrm{s}(k) - 1 = \frac{\dd\ln\mathcal{P}_\mathcal{R}}{\dd\ln k} \,, 
\end{equation}
which allows us to estimate the scalar spectral index in a given range of scales by computing the local logarithmic slope of the reconstructed PPS. Specifically, within each interval $[k_i,\,k_{i+1}]$, the spectral index can be obtained numerically, providing a scale-dependent measure of deviations from a simple power-law behavior.

In principle, this approach could also be extended to infer the running of the spectral index, $\alpha_\mathrm{s}$, or even higher-order runnings. However, in our case, the use of a piecewise linear interpolation makes higher derivatives trivially zero within each segment. Even if a higher-order interpolation, such as a cubic spline, were used, one should interpret higher derivatives with caution. The smoothness imposed by interpolation methods can introduce artificial features, potentially biasing the inferred values of the running of the scalar spectral index and beyond. 

\begin{figure}
    \centering
    \includegraphics[width=1\linewidth]{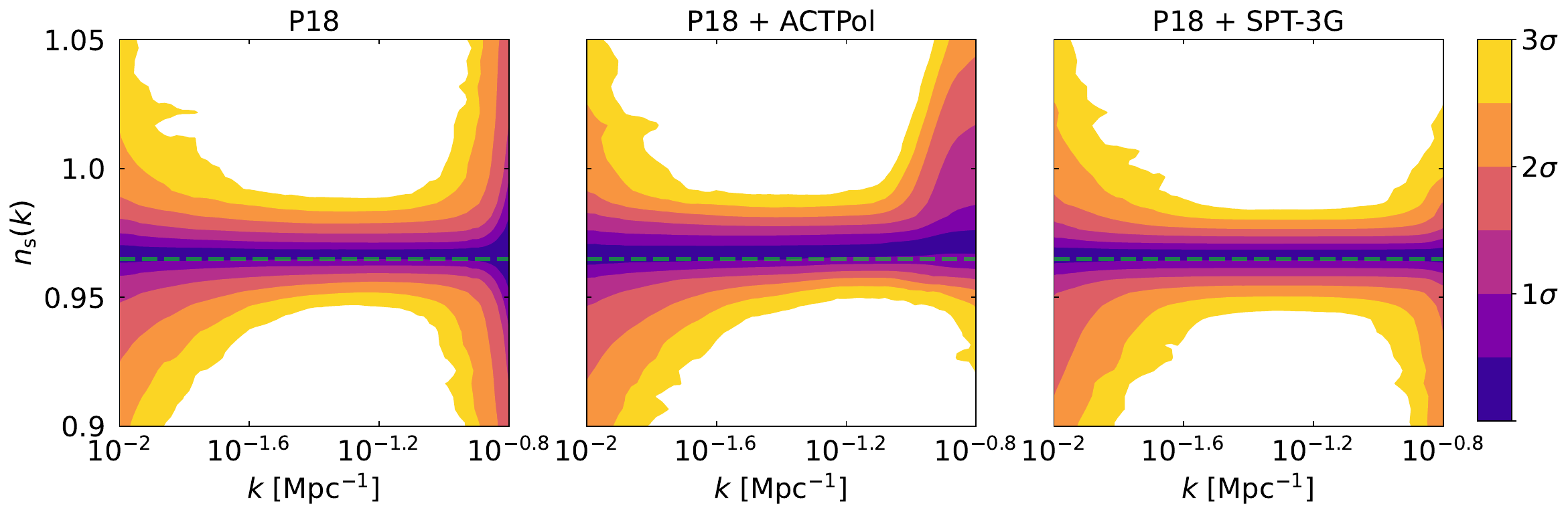}
    \caption{Reconstruction of the scalar spectral index $n_\mathrm{s}$ for $N=10$ obtained from the reconstructed scalar PPS analysed with log-uniform priors for the knots' positions. The green dashed line correspond to the \textit{Planck} PR3 mean value $n_\mathrm{s} = 0.9649$ \cite{Planck_cosmological_parameters_2020}.} 
    \label{fig:ns_10}
\end{figure}

In \cref{fig:ns_10}, we present the reconstructed scalar spectral index obtained using 10 knots logarithmic sampling reconstruction, illustrating how its scale dependence emerges from the reconstructed PPS.
The reconstructed $n_\mathrm{s}$ is in perfect agreement with the constant mean found using the $\Lambda$CDM model, assuming a power-law scalar PPS, corresponding to $n_\mathrm{s} = 0.9649 \pm 0.0041$ at 68\% CL for \textit{Planck} data \cite{Planck_cosmological_parameters_2020}; see \cref{fig:ns_10}.
In the region where the CMB has more constraining power, namely $k\in[10^{-2},10^{-0.8}]\, \mathrm{Mpc}^{-1}$, we find no deviation from a constant scalar spectral index from the 10-knot reconstruction results.

\subsection{Slow roll parameters}
\begin{figure}
    \centering
    \includegraphics[width=1\linewidth]{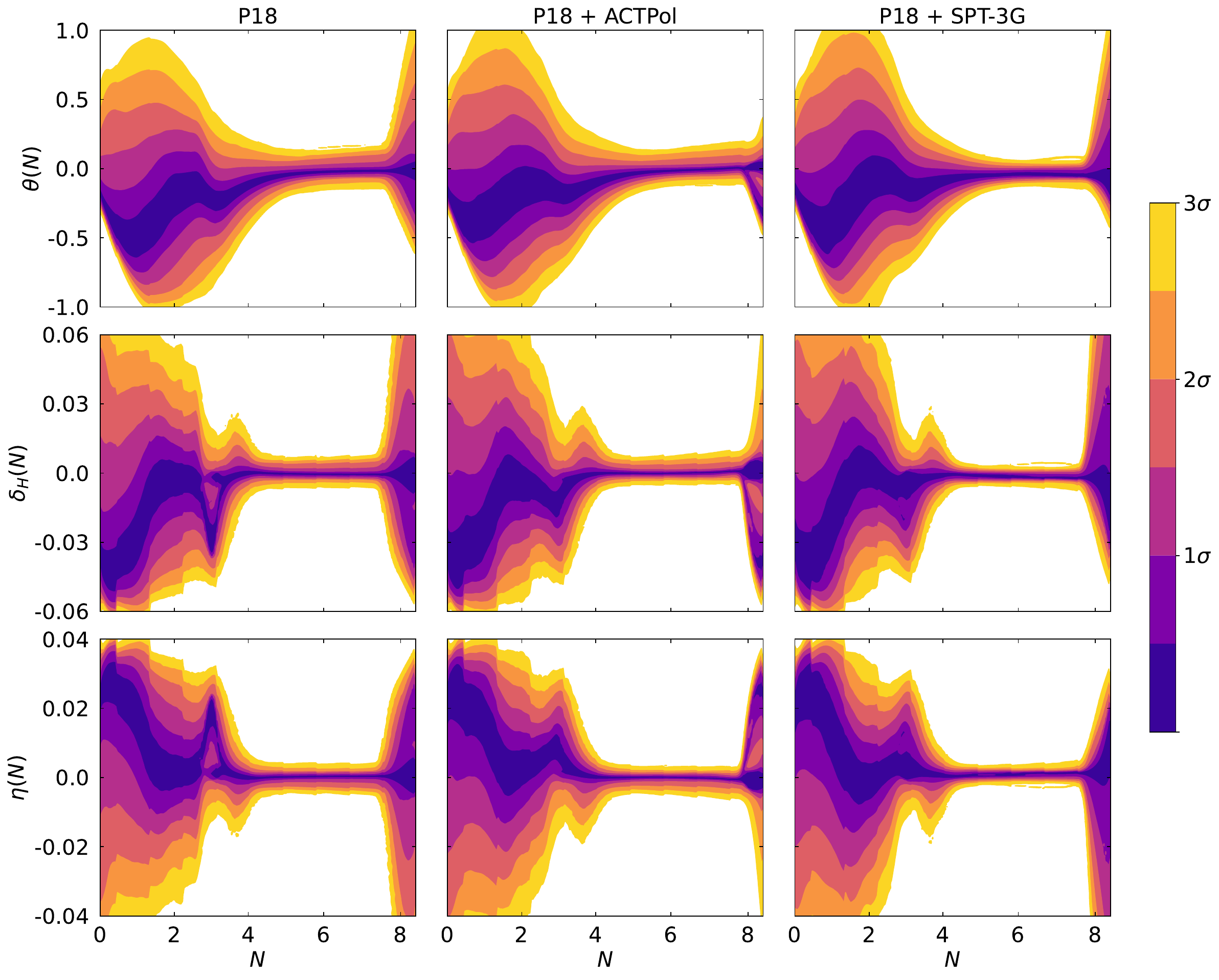}
    \caption{Reconstruction of slow roll parameters and $\theta$ from 10 knots logarithmic reconstruction. $\delta_H$ takes into account all the effect from slow-roll parameters excluding $\theta$, and $\eta$ is computed in the assumption of sudden variations of the background during inflation; $\theta$ is computed switching off effects from slow-roll and describes the evolution of the effective speed of sound $c_\mathrm{s}$.}
    \label{fig:slow_roll}
\end{figure}

In~\cref{fig:slow_roll}, we show the reconstruction obtained from the reconstructed PPS with $N=10$ adopting logarithmic sampling for the positions of the nodes of $\theta = 1-c_\mathrm{s}^2$, of the parameters $\delta_H$, defined as $\delta_H \simeq 3\epsilon+3\eta/2$, for a slow time-variation of the slow-roll parameters, and $\eta$ in case of sudden variations using~\cref{eqn:theta,eqn:deltaH,eqn:eta} 
respectively.
These plots can be interpreted as the behaviour of the parameters of the EFT of inflation at the time of horizon crossing. 
Perturbations in the PPS at different scales inherit the properties of these parameters at different times during inflation, giving a clear way of reading the results. We plot the parameters as function of the number of $e$-folds $N$. The relation between conformal time and the number of $e$-folds is 
\begin{equation}
    N=\ln\left(\frac{\tau_\text{in}}{\tau}\right)\,,
\end{equation}
where $\tau_*$ is the value of the conformal time during inflation at which a given scale $k_*$ crosses the horizon. We choose the reference scale $k_*=10^{-4}\,\text{Mpc}^{-1}$ corresponding to $\tau_*=-10^{4}$. This fixes $N=0$ when $k_*$ crosses the horizon.

We observe that the reconstructed parameters are particularly sensible to the features in the reconstructed PPS at large angular scales. Also we observe that in the regions where the reconstructed PPS is tightly constraint around a power-law form, no deviations compared to the single-field slow-roll results appear in the parameters.

\section{Conclusions} \label{sec:conclusions}
This study has focused on the reconstruction of the primordial power spectrum (PPS) of scalar fluctuations using a knot-spline approach, which provides a flexible and robust framework for exploring potential deviations from the standard power-law form assumed in a spatially-flat $\Lambda$CDM model. Similar methodologies have been employed in previous studies; see Refs.~\cite{Handley:2019fll,Planck_inflation_2018}. 

While no strong evidence for significant deviations from the power-law PPS was found, the method effectively identified regions of the spectrum where cosmic variance and instrumental noise reduce the ability to constrain features. This highlights the importance of using flexible models that account for these limitations in the data. In addition, our results show that the inclusion of more general PPS forms does not significantly affect the derived late-time cosmological parameters (see~\cref{app:parameters}), indicating the robustness of these parameters to changes in the primordial cosmology.

The knot-spline reconstruction approach also partially recovers oscillatory features in certain multipole regions, which could potentially be associated with a breakdown of slow-roll inflation. However, these features do not survive marginalization, suggesting that their significance is not strong enough to warrant a departure from the power-law baseline based on current data sets. Future observations, particularly with improved CMB polarization measurements and tighter constraints on reionisation, may provide the discriminatory power needed to validate or reject these potential features and further refine our understanding of the primordial universe.

By reconstructing the PPS, we extracted constraints on the scalar spectral index and the parameters governing the EFT of inflation derived in~\cref{sec:theory} for different regimes. In particular, we demonstrated that, at the scales where the analysed CMB experiments are most sensitive, these parameters remain consistent with the expected values within the $\Lambda$CDM model for $n_\mathrm{s}$ and with single-field slow-roll inflation for the effective inflationary parameters under consideration.

In principle, one could apply reconstruction methods to the tensor PPS, as done in \cite{Hiramatsu_2018}, but as discussed in \cite{Handley:2019fll}, with the state of the art measurements of the tensor to scalar ratio, the reconstruction of the tensor primordial power spectrum does not add any information to the scalar one, and we do not do that here.

On the other hand, large-scale structure (LSS) can be used to search for deviations from a scale-invariant PPS as primordial features \cite{Beutler:2019ojk,Ballardini:2022wzu,Mergulhao:2023ukp} and to further refine the reconstruction of the scalar PSS as shown in Refs.~\cite{Ravenni_2016,Martinez-Somonte:2023ckq}.
With the advent of LSS missions such as \textit{Euclid} \cite{Euclid:2024yrr} and DESI \cite{DESI:2024mwx}, it is crucial to ensure an accurate description of the LSS observables in the quasi nonlinear regime. In particular, when dealing with non-standard scalar PPS, special care must be taken to model scale-dependent effects and properly incorporate non-linear corrections, which can significantly affect the interpretation of the data; see Refs.~\cite{Vlah:2015zda,Vasudevan:2019ewf,Beutler:2019ojk,Ballardini:2019tuc,Li:2021jvz,Euclid:2023shr,Ballardini:2024dto}.

This work demonstrates the utility of non-parametric reconstructions in testing inflationary models, and paves the way for future investigations into the origin of primordial features and their implications for the physics of the early universe.

\acknowledgments
We thank Will Handley for useful comments on the manuscript. 
MB acknowledges financial support from the INFN InDark initiative and from the COSMOS network through the ASI (Italian Space Agency) Grants 2016-24-H.0, 2016-24-H.1-2018, 2020-9-HH.0 (participation in LiteBIRD phase A). AR acknowledges financial support from ASI-INFN Agreement No. 2019-19-HH.0. 
We acknowledge the use of computing facilities provided by the INFN theory group (I.S. InDark) at CINECA.

\appendix
\section{Cosmological parameter stability} \label{app:parameters}
\begin{figure}
    \centering
    \includegraphics[width=1\linewidth]{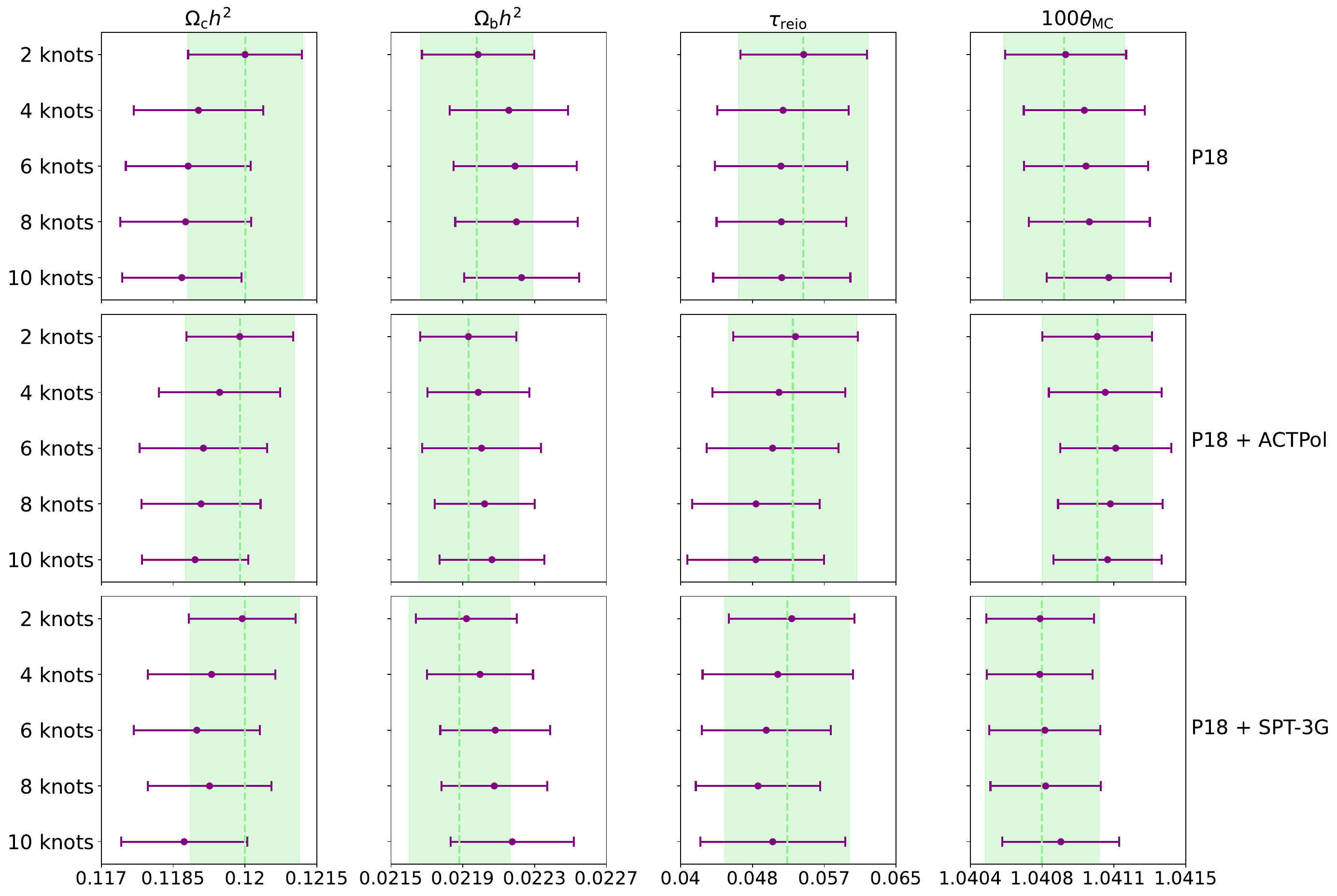}
    \caption{Mean values and 68\% CL of the posterior distribution for the standard cosmological parameters obtained from the reconstruction with knot's positions $k_i$ linearly sampled in the range $[10^{-4},10^{-0.3}]\, \mathrm{Mpc}^{-1}$ compared to the results obtained assuming the $\Lambda$CDM model with power-law PPS (green dashed linears and bands).}
    \label{fig:1d_lin}
\end{figure}

\begin{figure}
    \centering
    \includegraphics[width=1\linewidth]{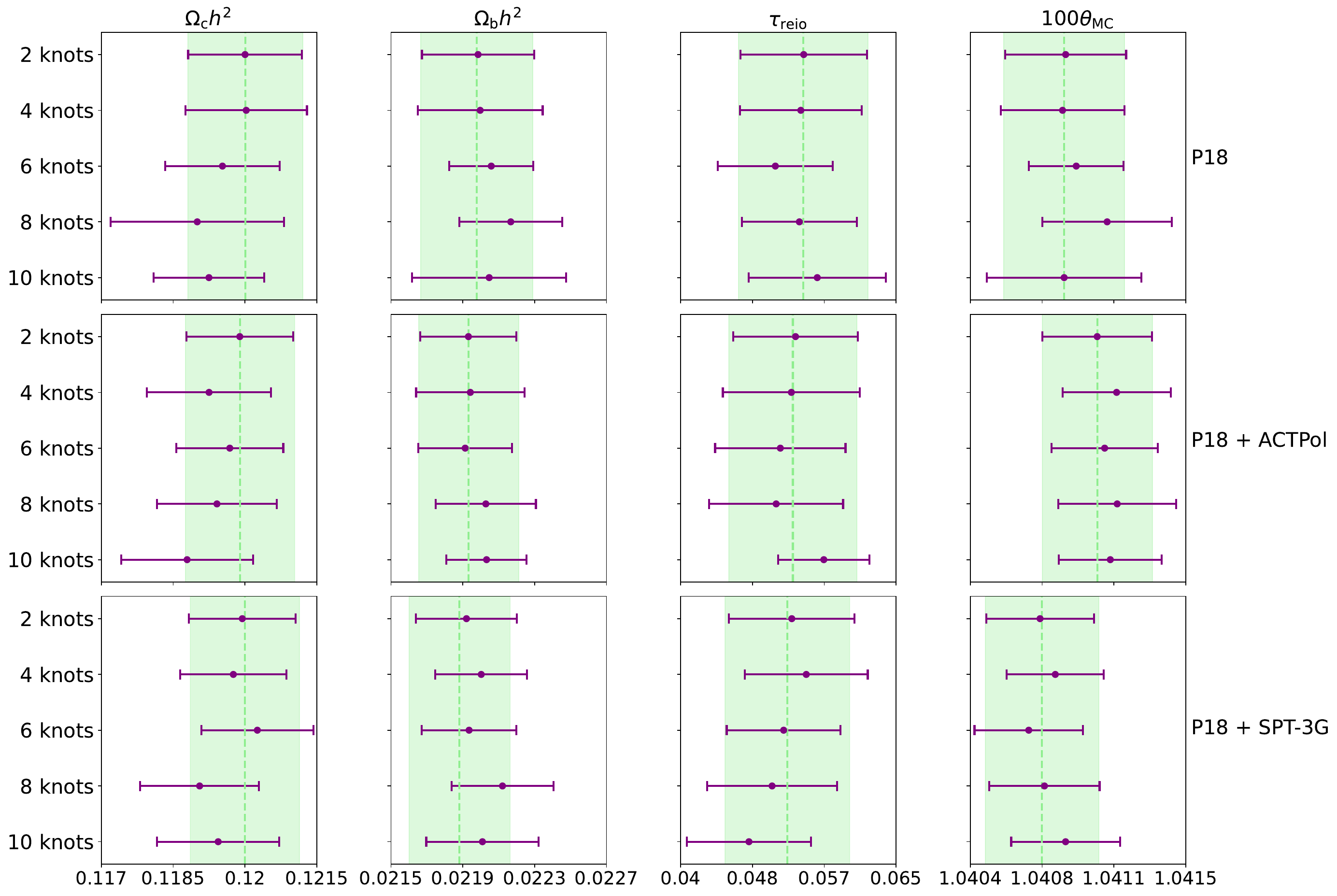}
    \caption{Same as \cref{fig:1d_lin}, but with knot's positions $k_i$ logarithmically sampled in the range $[10^{-4},10^{-0.3}]\, \mathrm{Mpc}^{-1}$.}
    \label{fig:1d_log}
\end{figure}

We present the mean values and the 68\% CL of the marginalised posterior distributions for the standard cosmological parameters varied during the reconstruction of the scalar PPS with logarithmic sampling in~\cref{fig:1d_log} and with linear sampling in~\cref{fig:1d_lin}.

For logarithmic sampling, we find that the derived cosmological parameters remain stable as the number of nodes $N$ increases, with no significant shift. In particular, all parameters are consistent at 68\% CL compared to the results obtained assuming a spatially flat $\Lambda$CDM model (green dashed lines and bands), indicating that introducing additional flexibility in the PPS does not significantly affect the constraints on the standard cosmological parameters.

For linear sampling, the parameters are still consistent with the $\Lambda$CDM model 68\% CL results. However, we observe a slight shift in $\Omega_\mathrm{c}h^2$ and $\Omega_\mathrm{b}h^2$ as the number of knots $N$ increases, suggesting a slight degeneracy between the reconstructed features at small scales in the PSS and the matter content. This behaviour is probably a consequence of the greater sensitivity of linear sampling to small-scale variations, which can induce small parameter shifts to compensate for local deviations in the reconstructed spectrum.

Overall, these results confirm the robustness of the cosmological parameter estimation and show that modifications to the PPS do not require significant deviations from the standard $\Lambda$CDM values to remain compatible with current CMB data.

\section{Comparison between \textit{Planck} PR3 and \textit{Planck} PR4} \label{app:npipe}
\begin{figure}
    \centering
    \includegraphics[width=1\linewidth]{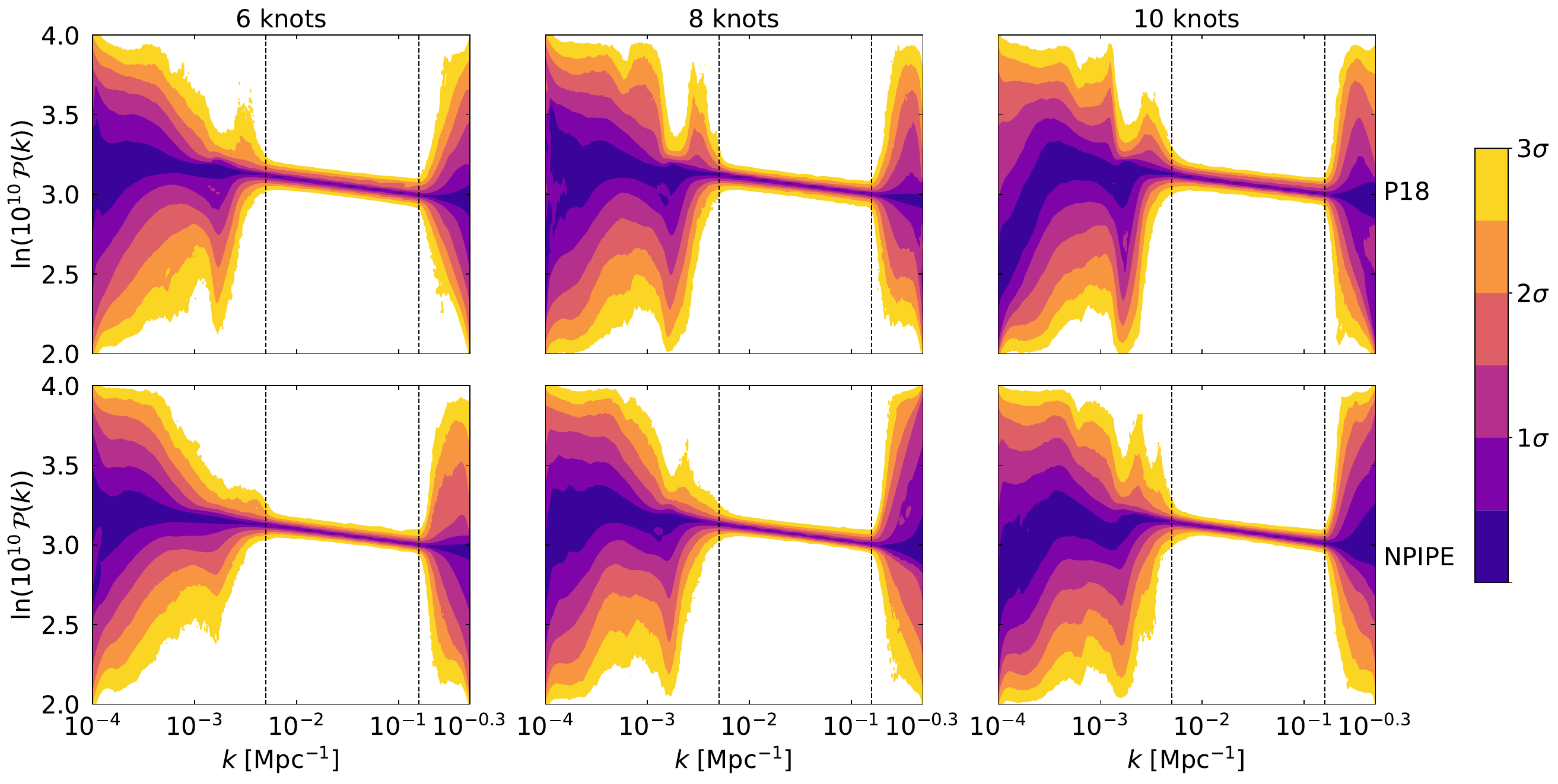}
    \caption{Same as~\cref{fig:spectra_log}, we compare the reconstructions obtained with P18 (upper panels) and the one obtained with NPIPE data sets and likelihoods (lower panels).}
    \label{fig:npipe}
\end{figure}

We present here a comparison of the reconstructed scalar PPS obtained using the \textit{Planck} PR3 data, as already presented in~\cref{sec:results}, and the results obtained using the updated reanalysis based on PR4 \texttt{NPIPE} maps and \texttt{SRroll2} low-$\ell$ polarisation data. The results in~\cref{fig:npipe} show no significant deviations in the reconstructed power spectrum. The results remain fully consistent within the expected uncertainties, reinforcing the robustness of the reconstruction across different data processing pipelines.

\section{Reconstruction results with $A_\text{L}$} \label{app:alens}
The \textit{Planck} PR3 data have consistently shown an anomalous smoothing of CMB acoustic peaks in temperature, milder using \textit{Planck} PR4 data \cite{camspec_npipe_2022}, often quantified by the phenomenological lensing amplitude $A_\mathrm{L}$ \cite{Calabrese_2008}, which exceeds the standard $\Lambda$CDM prediction of $A_\mathrm{L}=1$ at a significance of $2-3\sigma$. This anomaly produces oscillatory residuals in the CMB power spectrum, potentially mimicking oscillatory features in the PPS in coherent phase with the acoustic oscillations as discussed in Refs.~\cite{Hazra:2014jwa,Domenech:2019cyh,Planck_inflation_2018,Domenech:2020qay,Ballardini:2022vzh}.

To test for possible degeneracies, we repeat our PPS reconstruction varying $A_\text{L}$ over a uniform prior with range $[0,2]$. The results marginalising over $A_\mathrm{L}$ remain stable, with no significant changes in the extracted features, and continue to show a small deviation from unity. This confirms that the reconstruction does not absorb the effects of lensing anomalies, since our method naturally suppresses high-frequency oscillations.
Thus, any observed deviations from a power-law PPS are unlikely to be artefacts of the lensing anomaly, reinforcing the robustness of our approach.

\begin{figure}
    \centering
    \includegraphics[width=1\linewidth]{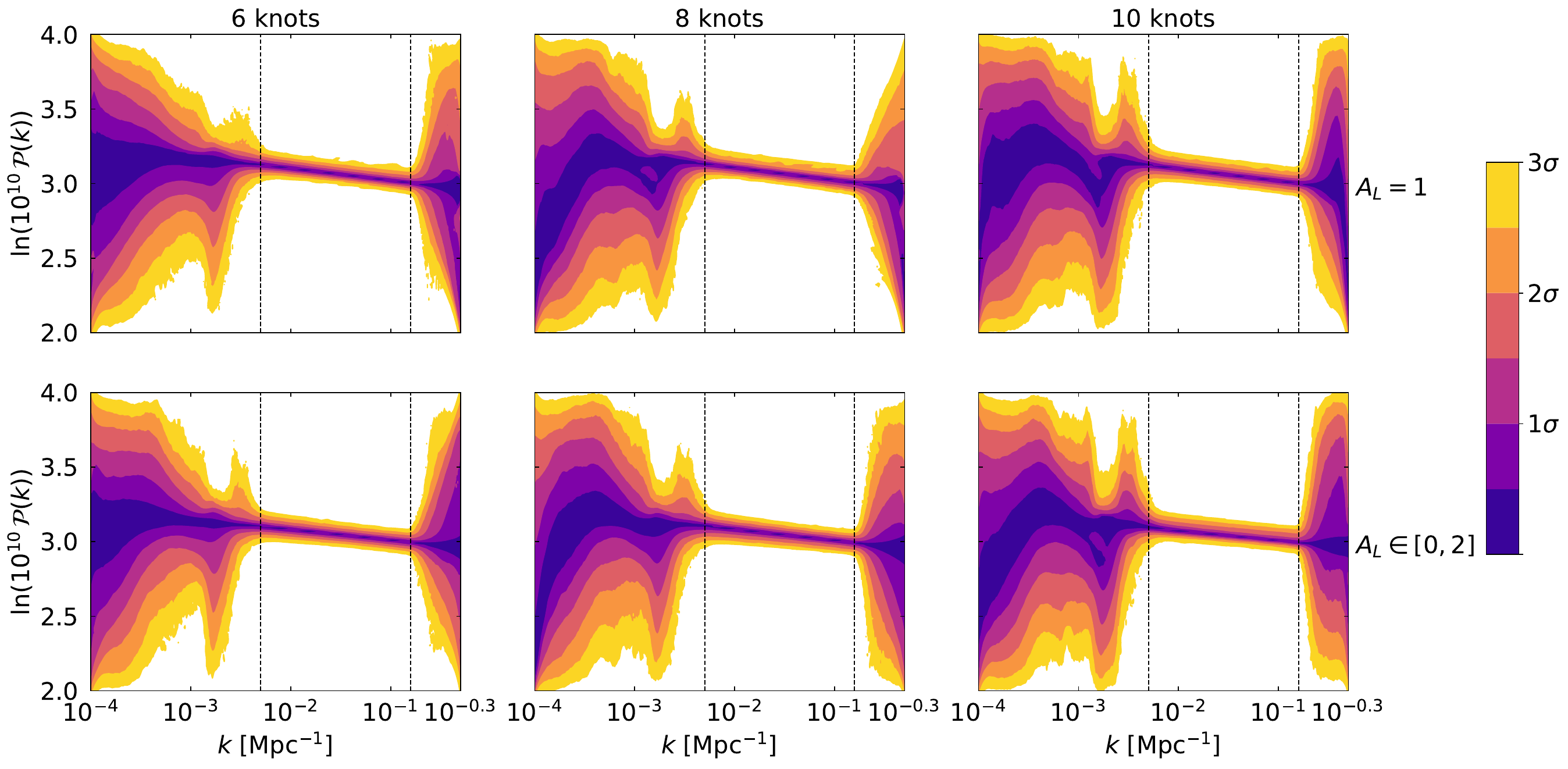}
    \caption{Same as~\cref{fig:spectra_log}, we compare the reconstructions obtained with the parameter $A_\text{L}$ fixed to unity (upper panels) with those obtained varying $A_\text{L}$ (lower panels) for P18 without using CMB lensing data.} 
    \label{fig:alens}
\end{figure}

The reconstructed spectra for 6, 8 and 10 knots are shown in~\cref{fig:alens}, where we use the \textit{Planck} PR3 likelihoods without the lensing one, which would reconcile $A_\text{L}$ with the standard prediction.

\begin{figure}
    \centering
    \includegraphics[width=1\linewidth]{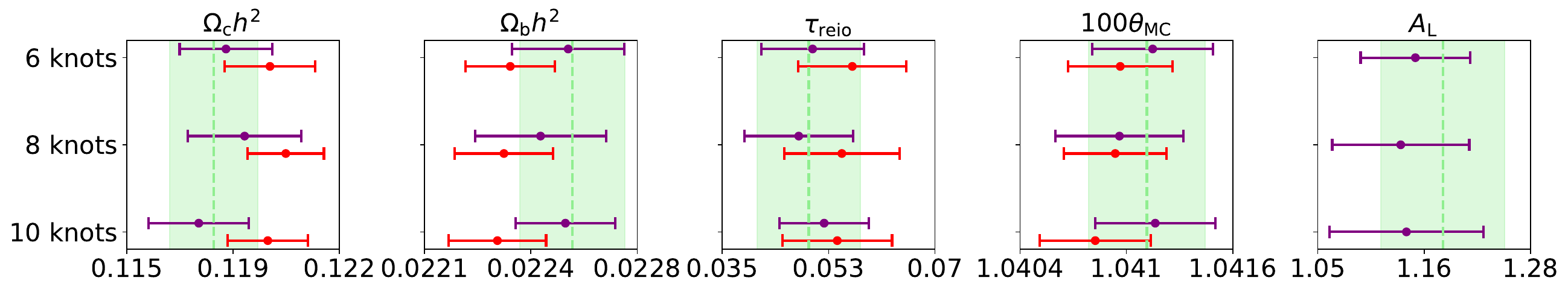}
    \caption{Marginalised means and 68\% CL of the posterior distributions of the cosmological parameters for reconstructions with free $A_\mathrm{L}$ (purple line) and with $A_\mathrm{L}=1$ (red line) for P18 without using CMB lensing data. The green dashed line corresponds to the means for the $\Lambda$CDM model with $A_\mathrm{L}\in[0,2]$ and the shaded area to their 68\% CL.}
    \label{fig:alens_1d}
\end{figure}

In~\cref{fig:alens_1d} we show the means and 68\% CL of the marginalised posterior distribution of the standard cosmological parameters plus $A_\mathrm{L}$ obtained from the reconstruction analysis and from the analysis assuming the $\Lambda$CDM model. 
All the parameters are compatible with the $\Lambda$CDM results within the 68\% CL showing that the reconstruction has no influence on the excess of smoothing in the \textit{Planck} PR3 temperature power spectrum.

\bibliographystyle{JHEP}
\bibliography{Biblio}

\end{document}